\documentclass[aps,preprint,nofootinbib,floatfix]{revtex4}
\usepackage{amssymb}
\usepackage{amsmath}
\usepackage{epsfig}

\input{tcilatex}

\begin{document}

\title{ZERO-NORM STATES AND SUPERSTRINGY SYMMETRIES}
\author{Jen-Chi Lee}
\address{Department of Electrophysics, National Chiao-Tung University,
Hsinchu, Taiwan, R.O.C.}
\email{jcclee@cc.nctu.edu.tw }
\date{\today }

\begin{abstract}
We identify spacetime symmetry charges of the bosonic sector of 10D
superstring theory from an infinite number of zero-norm states (ZNS) in the
old covariant first quantized string spectrum. We give evidences to support
this identification. These include supersymmetric sigma-model calculation,
2D super-Liouville theory calculation and, most importantly, three methods
of high-energy scattering amplitude calculations. These calculations
generalize the previous bosonic string calculations, which explicitly prove
Gross's conjectures in 1988 on high energy symmetry of string theory.
Moreover, we discover new high energy scattering amplitudes which are,
presumably, related to the high energy massive spacetime fermionic
scattering amplitudes in the R-sector of the theory.
\end{abstract}

\maketitle

\address{Department of Electrophysics, National Chiao Tung University,
Hsinchu 300, Taiwan}


\section{Introduction and Overview}

One of the fundamental issues of string theory is its spacetime symmetry. It
has long been believed that quantum string theory, which is UV finite and
contains an infinite number of states without any free parameter, consists
of a huge hidden spacetime symmetry. Historically, the first key progress to
understand symmetry of string theory is to study the high energy, fixed
angle behavior of string scattering amplitude \cite{GM,Gross,GrossManes} as
suggested by Gross in 1988. The second key idea to uncover the fundamental
symmetry of string theory was the identification of symmetry charges from an
infinite number of stringy zero-norm states (ZNS) with arbitrarily high
spins in the old covariant first quantized (OCFQ) string spectrum \cite{Lee}%
. In the context of $\sigma $-model approach of string theory, massive
inter-particle symmetries were calculated by using two types of ZNS. The
corresponding on-shell Ward identities on the scattering amplitudes were
constructed in \cite{JCLee,MassHeter}. On the other hand, ZNS were also
shown \cite{ChungLee} to carry the spacetime $\omega _{\infty }$ symmetry 
\cite{Winfinity} charges of 2D string theory \cite{2Dstring}. Incidentally,
it was also shown \cite{KaoLee,CLYang} that off-shell gauge transformations
of Witten string field theory \cite{Witten}, after imposing the no-ghost
condition, are identical to the on-shell stringy gauge symmetries generated
by two types of ZNS in the massive $\sigma $-model approach of string theory 
\cite{Lee}.

Recently high-energy Ward identities derived from the decoupling of ZNS of
the 26D open bosonic string, which combines the previous two key ideas of
probing stringy symmetry, were used to explicitly prove Gross's conjectures 
\cite{ChanLee1,ChanLee2,CHL,CHLTY1,CHLTY2,paperB,susy,Closed}. One utilizes
the decoupling of ZNS to obtain nonlinear relations or on-shell Ward
identities among string scattering amplitudes. In the high energy limit,
many simplications occur and one can derive linear relations among
high-energy scattering amplitudes of different string states at each fixed
mass levels. Moreover, these linear relations can be used to fix the ratios
among high energy scattering amplitudes of different string states at each
fixed mass level algebraically. This explicitly shows that there is only one
independent component of high-energy scattering amplitude at each mass
level. On the other hand, a saddle-point method was developed to calculate
the general formula of tree-level high-energy scattering amplitudes of four
arbitrary string states to verify the ratios calculated above. This general
formula expresses \textit{all} high-energy string scattering amplitudes in
terms of that of four tachyon as conjectured by Gross in 1988 \cite{Gross}.

In this report, we consider the high-energy scattering amplitudes for the
NS-sector of 10D open superstring theory \cite{susy}. Based on the
calculations of 26D bosonic open string \cite%
{ChanLee1,ChanLee2,CHL,CHLTY1,CHLTY2}, all the three independent
calculations of bosonic string, namely the decoupling of high-energy
zero-norm states (HZNS), the Virasoro constraints and the saddle-point
calculation, can be generalized to scattering amplitudes of string states
with polarizations on the scattering plane of superstring. All three methods
give the consistent results. In addition, we discover new leading order
high-energy scattering amplitudes, which are still proportional to the
previous ones, with polarizations \textit{orthogonal} to the scattering
plane. These scattering amplitudes are of subleading order in energy for the
case of 26D open bosonic string theory.

\section{\protect\bigskip Zero-norm state calculations}

In this section, we review the calculations of superstring symmetries from
ZNS without taking the high-energy limit. In the OCFQ spectrum of the
NS-sector of 10D open superstring theory, the solutions of physical states
conditions include positive-norm propagating states and two types of ZNS.
The latter are%
\begin{eqnarray}
\text{Type I} &:&G_{-\frac{1}{2}}\left\vert x\right\rangle ,\text{ where } 
\notag \\
G_{\frac{1}{2}}\left\vert x\right\rangle &=&G_{\frac{3}{2}}\left\vert
x\right\rangle =0,\text{ }L_{0}\left\vert x\right\rangle =0;  \label{1}
\end{eqnarray}%
\begin{eqnarray}
\text{Type II} &:&(G_{-\frac{3}{2}}+2G_{-\frac{1}{2}}L_{-1})\left\vert 
\widetilde{x}\right\rangle ,\text{ where }  \notag \\
G_{\frac{1}{2}}\left\vert \widetilde{x}\right\rangle &=&G_{\frac{3}{2}%
}\left\vert \widetilde{x}\right\rangle =0,\text{ }(L_{0}+1)\left\vert 
\widetilde{x}\right\rangle =0.  \label{2}
\end{eqnarray}%
While type I states have zero-norm at any space-time dimension, type II
states have zero-norm \emph{only} at D=10. In the supersymmetric $\sigma $%
-model approach of string theory, a spacetime symmetry transformation $%
\delta \Phi $ for a NS-NS bosonic background field $\Phi $ can be generated
by \cite{Super} 
\begin{equation}
\mathbf{T}_{\Phi }+\delta \mathbf{T}=\mathbf{T}_{\Phi +\delta \Phi },
\label{3}
\end{equation}%
where $\mathbf{T}_{\Phi }$ is the worldsheet superstress tensor with
background fields $\Phi $ and $\mathbf{T}_{\Phi +\delta \Phi }$ is the new
superstress tensor with new background fields $\Phi +\delta \Phi $. It can
be shown that for each ZNS in the NS-NS sector, one can construct a
superconformal deformation $\delta \mathbf{T}$ such that Eq.(3) is satisfied
to the first order of weak field approximation in the $\beta $ function
calculation. As an example, the massless ZNS solution of Eq.(1) can be used
to derive the on-shell gauge symmetries of massless graviton $h_{\mu \nu }$
and antisymmetric tensor $b_{\mu \nu }$ in the week field approximation.

Another evidence to support ZNS as the origin of symmetry charge was
demonstrated for the 2D super-Liouville theory. The spacetime symmetry of 2D
super-Liouville theory was known to be the $w_{\infty }$ algebra \cite{susyw}%
\begin{eqnarray}
&&\int \frac{d\mathbf{z}}{2\pi i}\mathbf{\psi }_{J_{1}M_{1}}^{+}(z)\mathbf{%
\psi }_{J_{2}M_{2}}^{+}(\mathbf{0})  \label{4} \\
&=&(J_{2}M_{1}-J_{1}M_{2})\mathbf{\psi }_{(J_{1}+J_{2}-1)(M_{1}+M_{2})}^{+}(%
\mathbf{0})  \notag
\end{eqnarray}%
generated by the discrete supersymmetric Polyakov states $\mathbf{\psi }%
_{JM}^{+}$. Alternatively, one can explicitly construct a set of discrete
ZNS $\mathbf{\Omega }_{JM}^{+}$ in the worldsheet supersymmetric form and
show that they form a $w_{\infty }$ algebra \cite{ChungLee}%
\begin{eqnarray}
&&\int \frac{d\mathbf{z}}{2\pi i}\mathbf{\Omega }_{J_{1}M_{1}}^{+}(\mathbf{z}%
)\mathbf{\Omega }_{J_{2}M_{2}}^{+}(\mathbf{0})  \label{5} \\
&=&(J_{2}M_{1}-J_{1}M_{2})\mathbf{\Omega }%
_{(J_{1}+J_{2}-1)(M_{1}+M_{2})}^{+}(\mathbf{0}).  \notag
\end{eqnarray}%
This seems to strongly suggest that ZNS are closely related to the spacetime
symmetry of string theory.

\section{High energy ZNS calculations}

Recently a further evidence to support ZNS as the spacetime symmetry charge
of string theory was obtained by taking the high-energy, fixed angle limit
of stringy Ward identities derived from the decoupling of ZNS on the
scattering amplitudes. The conjectures of Gross were then explicitly proved.
We first review the bosonic string case. At a fixed mass level $M^{2}=2(n-1)$
of 26D open bosonic string theory, it was shown that \cite{CHLTY1,CHLTY2} a
four-point function is at the leading order at high-energy limit only for
states of the following form%
\begin{eqnarray}
&&\left\vert n,2m,q\right\rangle  \label{6} \\
&\equiv &(\alpha _{-1}^{T})^{n-2m-2q}(\alpha _{-1}^{L})^{2m}(\alpha
_{-2}^{L})^{q}\left\vert 0,k\right\rangle .  \notag
\end{eqnarray}%
where $n\geqslant 2m+2q,m,q\geqslant 0.$Note that, in the high energy limit,
the scattering process becomes a plane scattering, and we have defined the
normalized polarization vectors of the second string state to be \cite%
{ChanLee1,ChanLee2} to be $e_{P}=\frac{1}{m_{2}}(E_{2},\mathrm{k}_{2},0)=%
\frac{k_{2}}{m_{2}},$ $e_{L}=\frac{1}{m_{2}}(\mathrm{k}_{2},E_{2},0)$ and $%
e_{T}$ $=(0,0,1)$ in the CM frame contained in the plane of scattering. By
using the decoupling of two types of bosonic string ZNS in the high energy
limit, an infinite linear relations among string scattering amplitudes at
mass level $M^{2}=2(n-1)$ can be derived \cite{CHLTY1,CHLTY2} 
\begin{eqnarray}
&&\mathcal{T}^{(n,2m,q)}  \label{7} \\
&=&\left( -\frac{1}{M}\right) ^{2m+q}\left( \frac{1}{2}\right) ^{m+q}(2m-1)!!%
\mathcal{T}^{(n,0,0)},  \notag
\end{eqnarray}%
where $\mathcal{T}^{(n,2m,q)}$ stands for the 4-point function with one
vertex in Eq.(6), and the other three any string vertex which we have
omitted their tensor index. Moreover, these linear relations can be used to
fix the ratios among high energy scattering amplitudes of different string
states at each mass level algebraically.

We now consider the superstring case \cite{susy}. We first consider
high-energy scattering amplitudes of string states with polarizations on the
scattering plane. It can be argued that there are four types of high-energy
scattering amplitudes for states in the NS-sector with even GSO parity.
These are%
\begin{align}
& \left\vert n,2m,q\right\rangle \otimes \left\vert b_{-\frac{1}{2}%
}^{T}\right\rangle  \label{8} \\
& \equiv (\alpha _{-1}^{T})^{n-2m-2q}(\alpha _{-1}^{L})^{2m}(\alpha
_{-2}^{L})^{q}(b_{-\frac{1}{2}}^{T})\left\vert 0,k\right\rangle ,  \notag \\
& \left\vert n,2m+1,q\right\rangle \otimes \left\vert b_{-\frac{1}{2}%
}^{L}\right\rangle  \label{9} \\
& \equiv (\alpha _{-1}^{T})^{n-2m-2q-1}(\alpha _{-1}^{L})^{2m+1}(\alpha
_{-2}^{L})^{q}(b_{-\frac{1}{2}}^{L})  \notag \\
& \left\vert 0,k\right\rangle ,  \notag \\
& \left\vert n,2m,q\right\rangle \otimes \left\vert b_{-\frac{3}{2}%
}^{L}\right\rangle  \label{10} \\
& \equiv (\alpha _{-1}^{T})^{n-2m-2q}(\alpha _{-1}^{L})^{2m}(\alpha
_{-2}^{L})^{q}(b_{-\frac{3}{2}}^{L})\left\vert 0,k\right\rangle ,  \notag \\
& \left\vert n,2m,q\right\rangle \otimes \left\vert b_{-\frac{1}{2}}^{T}b_{-%
\frac{1}{2}}^{L}b_{-\frac{3}{2}}^{L}\right\rangle  \label{11} \\
& \equiv (\alpha _{-1}^{T})^{n-2m-2q}(\alpha _{-1}^{L})^{2m}(\alpha
_{-2}^{L})^{q}(b_{-\frac{1}{2}}^{T})(b_{-\frac{1}{2}}^{L})  \notag \\
& (b_{-\frac{3}{2}}^{L})\left\vert 0,k\right\rangle .  \notag
\end{align}%
Note that the number of $\alpha _{-1}^{L}$ operator in Eq.(9) is odd. The
decoupling of ZNS in Eqs.(1) and (2) in the high energy limit implies the
ratios among scattering amplitudes%
\begin{align}
& \left\vert n+1,2m,q\right\rangle \otimes \left\vert b_{-\frac{1}{2}%
}^{T}\right\rangle  \label{12} \\
& \equiv \left( -\frac{1}{2M}\right) ^{m}\left( -\frac{1}{2M}\right) ^{q}%
\frac{\left( 2m-1\right) !!}{\left( -M\right) ^{m-1}}\left\vert
n,0,0\right\rangle  \notag \\
& \otimes \left\vert b_{-\frac{3}{2}}^{L}\right\rangle ,  \notag \\
& \left\vert n,2m+1,q\right\rangle \otimes \left\vert b_{-\frac{1}{2}%
}^{L}\right\rangle  \label{13} \\
& \equiv \left( -\frac{1}{2M}\right) ^{m}\left( -\frac{1}{2M}\right) ^{q}%
\frac{\left( 2m+1\right) !!}{\left( -M\right) ^{m+1}}\left\vert
n,0,0\right\rangle  \notag \\
& \otimes \left\vert b_{-\frac{3}{2}}^{L}\right\rangle ,  \notag \\
& \left\vert n,2m,q\right\rangle \otimes \left\vert b_{-\frac{3}{2}%
}^{L}\right\rangle  \label{14} \\
& \equiv \left( -\frac{1}{2M}\right) ^{m}\left( -\frac{1}{2M}\right) ^{q}%
\frac{\left( 2m-1\right) !!}{\left( -M\right) ^{m}}\left\vert
n,0,0\right\rangle  \notag \\
& \otimes \left\vert b_{-\frac{3}{2}}^{L}\right\rangle ,  \notag \\
& \left\vert n-1,2m,q-1\right\rangle \otimes \left\vert b_{-\frac{1}{2}%
}^{T}b_{-\frac{1}{2}}^{L}b_{-\frac{3}{2}}^{L}\right\rangle  \label{15} \\
& \equiv \left( -\frac{1}{2M}\right) ^{m}\left( -\frac{1}{2M}\right) ^{q}%
\frac{\left( 2m-1\right) !!}{\left( -M\right) ^{m}}\left\vert
n,0,0\right\rangle  \notag \\
& \otimes \left\vert b_{-\frac{3}{2}}^{L}\right\rangle .  \notag
\end{align}%
These ratios can be rederived from the method of Virasoro constraints and
the saddle-point calculation. Finally, it was discovered \cite{susy} that
many high-energy scattering amplitudes with polarizations \textit{orthogonal}
to the scattering plane are at the same order in energy as the previous ones
and are proportional to them. These scattering amplitudes are of subleading
order in energy for the case of 26D open bosonic string theory. The
existence of these new high-energy scattering amplitudes is due to the
worldsheet fermion exchange in the correlation functions and is, presumably,
related to the high energy massive fermionic scattering amplitudes in the
R-sector of the theory.

\section{Acknowledgments}

I thanks the collaborations of C.T. Chan and Y. Yang on calculations of high
energy superstring scatterings.

\end{document}